\documentclass[conference]{IEEEtran}
\ifCLASSINFOpdf
\else
   \usepackage{graphicx}
   \graphicspath{{../eps/}}
   \DeclareGraphicsExtensions{.eps}
\fi
\usepackage{fancyhdr}
\usepackage{amsmath,amssymb,amstext}
\usepackage{subeqnarray}
\usepackage{cite}
\usepackage{hhline}
\usepackage{bm}
\usepackage{soul}
\usepackage{slashbox}
\usepackage{times,amsmath,amsfonts}
\usepackage{subfigure}
\usepackage{amsmath,amssymb,bm}
\usepackage{graphicx,picture}
\usepackage{cite}
\usepackage{subfigure}
\usepackage{subeqnarray}
\usepackage{cases}
\usepackage{color}
\usepackage{overpic}
\usepackage{multirow}
\usepackage{booktabs}
\usepackage{epstopdf}
\usepackage[noend]{algpseudocode}
\usepackage{algorithmicx,algorithm}

\definecolor{r}{rgb}{1,0,0}
\definecolor{b}{rgb}{0,0,1}
\definecolor{k}{rgb}{0,1,1}

\usepackage{upgreek}

\newcounter{saveeqn}%

\allowdisplaybreaks
\DeclareMathSymbol{\Phi}{\mathord}{letters}{8}

\begin{document}
\title{Integrated Sensing and Communication with Delay Alignment Modulation}

\author{
\IEEEauthorblockN{$\text{Zhiqiang~Xiao}^*$ and $\text{Yong~Zeng}^*\ddagger$}

\IEEEauthorblockA{*National Mobile Communications Research Laboratory, Southeast University, Nanjing 210096, China\\
$\ddagger$Purple Mountain Laboratories, Nanjing 211111, China\\
Email: zhiqiang\_xiao@seu.edu.cn, yong\_zeng@seu.edu.cn
}
}
\maketitle

\begin{abstract}
\textsl{\textbf{Delay alignment modulation}} (DAM) has been recently proposed to enable inter-symbol interference (ISI)-free single-carrier (SC) communication without relying on sophisticated channel equalization.
The key idea of DAM is to pre-introduce deliberate symbol delays at the transmitter side, so that all multi-path signal components may arrive at the receiver simultaneously and be superimposed constructively, rather than causing the detrimental ISI.
Compared to the classic orthogonal frequency division multiplexing (OFDM) transmission, DAM has several appealing advantages, including low peak-to-average-power ratio (PAPR) and high tolerance for Doppler frequency shift, which makes DAM also appealing for radar sensing.
Therefore, in this paper, DAM is investigated for the emerging integrated sensing and communication (ISAC) setup.
We first derive the output signal-to-noise ratios (SNRs) for ISI-free communication and radar sensing, respectively, and then propose an efficient beamforming design for DAM-ISAC to maximize the communication SNR while guaranteeing the sensing performance.
The comparison analysis of DAM versus OFDM for ISAC is developed, and it is revealed that DAM enables higher sensing SNR and larger Doppler frequency estimation.
Simulation results are provided to show the great potential of DAM for ISAC.
\end{abstract}

\section{Introduction}
Integrated sensing and communication (ISAC) is an emerging new area that attracts increasing attention recently, which aims to efficiently utilize the precious radio resources and hardware for dual purposes of sensing and communication \cite{zheng2019radar,liu2020joint,wang2021SNR}.
To practically realize ISAC for future wireless networks, one of the critical issues is to find the most suitable waveform or modulation technique to fully exploit the performance gain of ISAC for both communication and sensing \cite{xiao2021full}.

To that end, early research efforts were mainly devoted to  radar-centric methods \cite{hassanien2016signaling}, where the communication symbols are usually embedded into the radar waveforms to achieve both sensing and communication purposes.
However, such techniques suffer from the extremely low communication rate.
Alternatively, ISAC can also be implemented based on communication-centric waveforms, like orthogonal frequency division multiplexing (OFDM) waveform \cite{sturm2011waveform}.
On one hand, OFDM is a mature communication technology, with successful applications in the fourth- (4G) and fifth-generation (5G) cellular networks, as well as wireless local area network (WLAN) \cite{heath2018foundations}.
On the other hand, as a multi-carrier technique, OFDM-based radar sensing may decouple the delay and Doppler frequency processing by simply applying the Fast Fourier Transform (FFT) and inverse-FFT (IFFT) \cite{sturm2011waveform}.
However, OFDM-based waveform also has several practical drawbacks for ISAC.
First, OFDM is known to suffer from high peak-to-average-power ratio (PAPR), which is proportional to the number of the subcarriers \cite{tse2005fundamentals}.
Second, the orthogonality across subcarriers is difficult to be guaranteed in high-mobility scenarios, which may lead to a severe performance degradation for both communication and sensing.
Recently, a new multi-carrier technique termed {\it orthogonal time frequency space} (OTFS) modulation has been proposed \cite{hadani2017orthogonal}, which modulates data symbols in the delay-Doppler domain, so that each data symbol is spread over the entire time-frequency diversity.
In \cite{raviteja2019orthogonal,gaudio2019effectiveness,yuan2021integrated}, OTFS were studied for radar systems, which showed that its superior performance than OFDM in high-mobility scenarios.
However, OTFS-based ISAC requires sophisticated signal processing for transmission and reception, and its study is still in the infancy stage.

In this paper, we investigate ISAC with a recently proposed transmission technology, termed {\it delay alignment modulation} (DAM) \cite{dam2021lu}.
By exploiting the high spatial resolution with large antenna arrays and multi-path sparsity of millimeter wave (mmWave) or  Terahertz wireless channels, the key idea of DAM is to pre-introduce deliberate symbol delays at the transmitter side, so that all multi-path signal components may arrive at the receiver simultaneously and be superimposed constructively, rather than causing the detrimental inter-symbol interference (ISI) \cite{dam2021lu}.
As a result, DAM enables ISI-free communication by low complexity single-carrier (SC) transmission without relying on sophisticated channel equalization.
Therefore, compared to OFDM, DAM can resolve the practical issues like high PAPR and strict requirements on the signal orthogonality in time-frequency domain.
Such appealing features also render DAM attractive for radar sensing.
Specifically, with reduced PAPR, DAM may allow higher average transmit power than OFDM before those nonlinear devices like power amplifiers get saturated.
Moreover, different from OFDM that typically tolerates the Doppler frequency shift only about 10\% of the subcarrier spacing, and hence with limited velocity estimation capability, the tolerable Doppler frequency shift is significantly enhanced for DAM.
In addition, different from those classic SC radar waveforms for ISAC, like frequency modulated continuous waveform (FMCW) or ultra-wideband (UWB) waveform, DAM can achieve superior communication spectral efficiency without channel equalization.

In this paper, the performance of DAM is studied for a basic multiple-input single-output (MISO) ISAC system, where a monostatic multi-antenna ISAC node wishes to communicate with an user equipment (UE) while simultaneously sensing a target.
With the novel DAM transmission, both the communication signal-to-noise ratio (SNR) and the sensing out SNR are derived, both of which depend on the per-path transmit beamforming vectors.
Therefore, a beamforming optimization problem is formulated for DAM-ISAC, which aims to maximize the communication SNR while guaranteeing that the sensing output SNR is no smaller than a desired threshold.
Furthermore, the comparison analysis of DAM versus OFDM is developed, in terms of the sensing ambiguity, resolution, and PAPR, which showed that DAM has higher sensing output SNR and can estimate larger Doppler frequency.
Finally, simulation results are provided to demonstrate the great potential of DAM for ISAC.

\section{System Model And Delay alignment modulation}
\begin{figure} 
  \setlength{\abovecaptionskip}{-0.1cm}
  \setlength{\belowcaptionskip}{-0.3cm}
  \centering
  \includegraphics[width=0.38\textwidth]{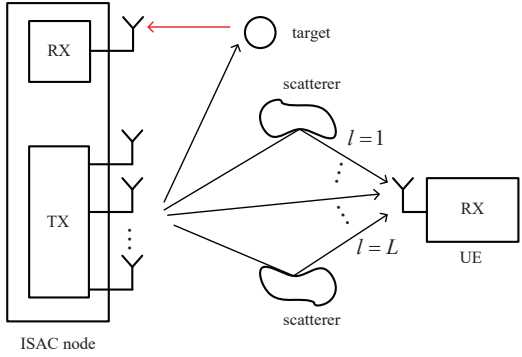}
  \caption{A monostatic ISAC system, where a multi-antenna ISAC node wishes to simultaneously communicate with an UE and senses a target.}\label{system}\vspace{-0.6cm}
\end{figure}

As shown in Fig.~\ref{system}, we consider a monostatic ISAC system, where an ISAC node equipped with $M$ transmit antennas and one receive antenna wishes to simultaneously communicate with a single-antenna UE and senses a radar target.
Denote by $B$ the total bandwidth available for the system.
For wideband multi-path communication environment, the MISO channel between the ISAC transmitter and the communication UE can be modelled as
\vspace{-0.2cm}
\begin{equation}\label{comm_model}
\vspace{-0.2cm}
\mathbf{h}^H_c[n] = \sum\nolimits_{l=1}^L \mathbf{h}_l^H\delta[n-n_l],
\end{equation}
where $L$ denotes the number of temporal resolvable multi-paths with delay resolution of $T_s=1/B$; $\mathbf{h}_l\in\mathbb{C}^{M\times1}$ and $n_l$ are the complex channel coefficients vector and symbol delay of the $l$th path, respectively.

We assume the clear  line-of-sight (LoS) path between ISAC node and the sensing target.
Furthermore, the self-interference leaked from the transmit antennas of the ISAC node to its receive antenna is assumed to be effectively suppressed.
Thus, the round-trip channel for radar sensing is given by
\vspace{-0.1cm}
\begin{equation}\label{radar_model}
\vspace{-0.1cm}
\mathbf{h}_s^H[n] = \alpha\delta[n-n_s]e^{j2\pi f_d nT_s}\mathbf{a}^H(\theta),
\end{equation}
where $\alpha$ denotes the complex channel coefficient of the round-trip channel;
$n_s=\mathrm{round}(\tau B)$ is the two-way propagation delay in terms of the number of symbols, with $\tau$ being the two-way propagation delay in second and $1/B$ being the delay resolution, and $\mathrm{round}(\cdot)$ denotes rounding to the nearest integer;
$f_d=\frac{2v_d}{\lambda}$ is the Doppler frequency caused by the motion of the target with the radial velocity $v_d$, and $\lambda$ is carrier wavelength;
$\mathbf{a}(\theta)$ is the transmit steering vector, with $\theta$ denoting the direction of the sensing target.

With DAM, the transmitted signal by the $M$-element transmit array is given by \cite{dam2021lu}
\vspace{-0.2cm}
\begin{equation}\label{dam_tx}
\vspace{-0.2cm}
\mathbf{x}[n]=\sum\nolimits_{l=1}^{L}\mathbf{f}_ls[n-\kappa_l],
\end{equation}
where $s[n]$ denotes the independent and identically distributed (i.i.d) information-bearing symbols with unit power $\mathbb{E}[|s[n]|^2]=1$;
$\kappa_l$ is the pre-introduced delay for the $l$th path, with $\kappa_l\neq\kappa_{l'},\forall l\neq l'$, and $\mathbf{f}_l\in\mathbb{C}^{M\times1}$ denotes the per-path transmit beamforming vector for the $l$th path.
The transmit power of $\mathbf{x}[n]$ is
\vspace{-0.2cm}
\begin{equation}\label{power_const}
\vspace{-0.2cm}
\mathbb{E}\left[\left\|\mathbf{x}[n]\right\|^2\right] = \sum_{l=1}^L\mathbb{E}\left[\left\|\mathbf{f}_ls[n-\kappa_l\right\|^2\right] = \sum_{l=1}^L\left\|\mathbf{f}_l\right\|^2\le P,
\end{equation}
where $P$ denotes the average transmit power constraint.
For the communication UE with channel impulse response given by \eqref{comm_model}, the received signal is
\vspace{-0.2cm}
\begin{equation}\label{dam_rx}
\vspace{-0.2cm}
\begin{aligned}
y_c[n] &= \sum\nolimits_{l=1}^L\mathbf{h}_l^H\mathbf{x}[n-n_l] + z[n]\\
& = \sum\limits_{l=1}^{L}\sum\limits_{l'=1}^{L}\mathbf{h}_l^H\mathbf{f}_{l'}s[n-\kappa_{l'}-n_l]+z[n],
\end{aligned}
\end{equation}
where $z[n]\sim\mathcal{CN}(0,\sigma^2)$ is the additive white Gaussian noise (AWGN).
By letting $\kappa_l=n_{\max}-n_l, \forall l=1,\cdots,L$, with $n_{\max}=\max\limits_{1\le l\le L}n_l$, \eqref{dam_rx} can be rewritten as
\vspace{-0.2cm}
\begin{equation}\label{dam_opt}
\vspace{-0.2cm}
\begin{aligned}
&y_c[n] = \left(\sum\nolimits_{l=1}^{L}\mathbf{h}_l^H\mathbf{f}_l\right)s[n-n_{\max}]\\
&+\sum\nolimits_{l=1}^{L}\sum\nolimits_{l'\neq l}^{L}\mathbf{h}_l^H\mathbf{f}_{l'}s[n-n_{\max}+n_{l'}-n_l]+z[n],
\end{aligned}
\end{equation}
where all multi-paths are aligned with the maximum delay $n_{\max}$ and contribute to the desired signal in the first term of \eqref{dam_opt}, while the residual ISI in the second term of \eqref{dam_opt} can be effectively suppressed by applying appropriate per-path transmit beamforming.
For example, if $\left\{\mathbf{f}_l\right\}_{l=1}^L$ are designed so that $\mathbf{h}_l^H\mathbf{f}_{l'}=0,\forall l\neq l'$, the received DAM signal in \eqref{dam_opt} reduces to
\vspace{-0.2cm}
\begin{equation}\label{ISI-ZF}
\vspace{-0.2cm}
y_c[n] = \left(\sum\nolimits_{l=1}^{L}\mathbf{h}_l^H\mathbf{f}_l\right)s[n-n_{\max}] + z[n],
\end{equation}
which is referred to as the ISI zero-forcing (ISI-ZF) beamforming \cite{dam2021lu} and it can be achieved as long as $M\ge L$, i.e., when the channel is sparse and/or the antenna dimension is large.
Note that the original multi-path channel with ISI in \eqref{dam_opt} has been transformed into an ISI-free AWGN channel in \eqref{ISI-ZF} with a single delay $n_{\max}$.
Thus, the received SNR of \eqref{ISI-ZF} is
\vspace{-0.2cm}
\begin{equation}\label{comm_snr}
\vspace{-0.2cm}
\gamma_c = {\left|\sum\nolimits_{l=1}^L\mathbf{h}_l^H\mathbf{f}_l\right|^2}/\sigma^2.
\end{equation}

\section{Performance Analysis and Beamforming for DAM-ISAC Optimization}\label{performance analysis}
\subsection{DAM Sensing Performance Analysis}
To evaluate the sensing performance by the DAM signal in \eqref{dam_tx},
let $N$ be the number of symbol durations for a coherent processing interval (CPI) for radar sensing. Then the transmitted signals in \eqref{dam_tx} for a block of $N$ symbol durations can be expressed as
\begin{equation}\label{dam_tx2}
\bar{\mathbf{X}}[n] = \big[\mathbf{x}[n-(N-1)],\cdots,\mathbf{x}[n]\big]\in\mathbb{C}^{M\times N}.
\end{equation}
Denote by
$\mathbf{F}\triangleq[\mathbf{f}_1,\cdots,\mathbf{f}_L]\in\mathbb{C}^{M\times L}$ the transmit beamforming matrix, and $\bar{\mathbf{S}}[n]\triangleq\big[\mathbf{s}[n-\kappa_1],\cdots,\mathbf{s}[n-\kappa_L]\big]^T\in\mathbb{C}^{L\times N}$ the transmitted symbols consisting of $\mathbf{s}[n]=\big[s[n-(N-1)],\cdots,s[n]\big]^T\in\mathbb{C}^{N\times 1}$ with different pre-introduced delays, $\kappa_1, \cdots, \kappa_L$.
Therefore, based on \eqref{dam_tx}, the transmitted  DAM signal for a block of $N$ symbol durations can be rewritten as
$
\bar{\mathbf{X}}[n] = \mathbf{F}\bar{\mathbf{S}}[n].
$
As a result, for a sensing target with channel given in \eqref{radar_model}, the received echo signal is
\vspace{-0.2cm}
\begin{equation}\label{dam_echo}
\vspace{-0.2cm}
\begin{aligned}
\bar{\mathbf{y}}_s^H[n] &= \alpha\mathbf{a}^H(\theta)\bar{\mathbf{X}}[n-n_s]\mathrm{diag}(\mathbf{d}[f_d]) + \mathbf{z}^H \\
&=\alpha\mathbf{a}^H(\theta)\mathbf{F}\bar{\mathbf{S}}[n-n_s]\mathrm{diag}(\mathbf{d}[f_d]) + \mathbf{z}^H,
\end{aligned}
\end{equation}
where $\mathbf{d}[f_d]\triangleq[e^{j2\pi f_d(n-(N-1))T_s},\cdots,e^{j2\pi f_dnT_s}]^T$ is the Doppler frequency vector, and $\mathbf{z}\in\mathbb{C}^{N\times 1}$ is the i.i.d AWGN vector with $\mathbb{E}[\mathbf{z}\mathbf{z}^H]=\sigma^2\mathbf{I}_N$.

Note that with the monostatic architecture, the transmitted signals $\bar{\mathbf{X}}[n]$ are known for the ISAC receiver.
Furthermore, as we mainly focus on the delay-Doppler sensing for the considered setup, we assume that the target direction $\theta$ is known, say via beam scanning in the searching mode.
Therefore, with the received signal vector $\bar{\mathbf{y}}_s[n]\in\mathbb{C}^{N\times 1}$ in \eqref{dam_echo}, matched filter can be constructed for each delay-Doppler bin parameterized by $(n_p,f_q)$ as
\begin{equation}
\begin{aligned}
&\mathbf{h}^H(n_p,f_q) = \frac{\mathbf{a}^H(\theta)\mathbf{F}\bar{\mathbf{S}}[n-n_p]\mathrm{diag}(\mathbf{d}[f_q])}{\left\|\mathbf{a}^H(\theta)\mathbf{F}\bar{\mathbf{S}}[n-n_p]\mathrm{diag}(\mathbf{d}[f_q])\right\|},
\\
&p = 0,\cdots,P-1, q = 0,\cdots, Q-1,
\end{aligned}
\end{equation}
where $P$ and $Q$ denote the number of bins in the delay and Doppler frequency interval of interest, respectively, with delay resolution of $1/B$ and the Doppler frequency resolution of $1/(NT_s)$;
$\mathbf{d}[f_q]\triangleq\left[e^{j2\pi f_q(n-(N-1))T_s},\cdots,e^{j2\pi f_qnT_s}\right]^T$ is defined as the Doppler frequency estimator, with $f_q$ denoting the Doppler frequency corresponding to the $q$th Doppler bin.
After matched filtering (MF), the resulting output is given by
\begin{equation}\label{mf}
\begin{aligned}
{r}(n_p,f_q) &= \mathbf{y}_s^H[n]\mathbf{h}(n_p,f_q) \\
&=\alpha\mathbf{a}^H(\theta)\mathbf{F}\bar{\mathbf{S}}[n-n_s]\mathrm{diag}(\mathbf{d})\\
&\quad\times\frac{\mathrm{diag}^H(\mathbf{d}[f_q])\bar{\mathbf{S}}^H[n-n_p]\mathbf{F}^H\mathbf{a}(\theta)}{\left\|\mathbf{a}^H(\theta)\mathbf{F}\bar{\mathbf{S}}[n-n_p]\mathrm{diag}(\mathbf{d}[f_q])\right\|}+\hat{z},\\
\end{aligned}
\end{equation}
where $\hat{z}\triangleq\frac{\mathbf{z}^H\mathrm{diag}^H(\mathbf{d}[f_q])\bar{\mathbf{S}}^H[n-n_p]\mathbf{F}^H\mathbf{a}(\theta)}{\left\|\mathbf{a}^H(\theta)\mathbf{F}\bar{\mathbf{S}}[n-n_p]\mathrm{diag}(\mathbf{d}[f_q])\right\|}$ is the resulting AWGN after MF, with distribution $\hat{z}\sim\mathcal{CN}(0,\sigma^2)$.
For the Doppler bin where the target actually lies, we have $f_q=f_d$ and $\mathrm{diag}(\mathbf{d}[f_d])\mathrm{diag}^H(\mathbf{d}[f_q])=\mathbf{I}_N$.
Thus \eqref{mf} reduces to
\begin{equation}\label{mf2}
\begin{aligned}
{r}(n_p)&=\frac{\alpha\mathbf{a}^H(\theta)\mathbf{F}\bar{\mathbf{S}}[n-n_s]\bar{\mathbf{S}}^H[n-n_p]\mathbf{F}^H\mathbf{a}(\theta)}{\left\|\mathbf{a}^H(\theta)\mathbf{F}\bar{\mathbf{S}}[n-n_p]\mathrm{diag}(\mathbf{d}[f_q])\right\|}+\hat{{z}}\\
&=\frac{\alpha\mathbf{a}^H(\theta)\mathbf{F}\bm{\Lambda}(n_p,n_s)\mathbf{F}^H\mathbf{a}(\theta)}{\sqrt{\mathbf{a}^H(\theta)\mathbf{F}\bm{\Lambda}(n_p,n_p)\mathbf{F}^H\mathbf{a}(\theta)}}+\hat{z},
\end{aligned}
\end{equation}
where $\bm{\Lambda}(n_p,n_s)\triangleq\bar{\mathbf{S}}[n-n_s]\bar{\mathbf{S}}^H[n-n_p]\in\mathbb{C}^{L\times L}$ is defined as the correlation matrix, which is a function of the delay bin $n_p$ and the ground truth delay $n_s$.
Moreover, for the delay bin where the target actually lies, i.e., $n_p = n_s$, \eqref{mf2} reduces to
\vspace{-0.2cm}
\begin{equation}\label{pmf}
\vspace{-0.2cm}
r(n_s) = \alpha\sqrt{\mathbf{a}^H(\theta)\mathbf{F}\bm{\Lambda}(n_s,n_s)\mathbf{F}^H\mathbf{a}(\theta)} + \hat{z}.
\end{equation}
Note that the element of the correlation matrix $\bm{\Lambda}(n_p,n_s)$ at the $i$th row and $j$th column is given by
\vspace{-0.2cm}
\begin{equation}\label{element}
\vspace{-0.2cm}
\left[\bm{\Lambda}(n_p,n_s)\right]_{i,j} = \mathbf{s}^H[n-\kappa_j-n_s]\mathbf{s}[n-\kappa_i-n_p], 1\le i,j\le L.
\end{equation}
Furthermore, since $s[n]$ are i.i.d communication symbols with normalized power 1, when $N$ is large, we have
\vspace{-0.2cm}
\begin{equation}
\vspace{-0.2cm}
\mathbf{s}^H[n_1]\mathbf{s}[n_2] \rightarrow \mathbb{E}[\mathbf{s}^H[n_1]\mathbf{s}[n_2]]= \left\{
\begin{aligned}
N, && n_1 = n_2,\\
0, && n_1\neq n_2.
\end{aligned}
\right.
\end{equation}
Therefore, it follows from \eqref{element} that
\vspace{-0.2cm}
\begin{equation}\label{element2}
\vspace{-0.2cm}
\left[\bm{\Lambda}(n_p,n_s)\right]_{i,j}= \left\{
\begin{aligned}
N, && \kappa_i+n_p = \kappa_j+n_s\\
0, && \kappa_i+n_p \neq \kappa_j+n_s
\end{aligned}, 1\le i,j\le L.
\right.
\end{equation}
When $n_p = n_s$, \eqref{element2}  reduces to
\vspace{-0.2cm}
\begin{equation}
\vspace{-0.2cm}
\left[\bm{\Lambda}(n_s,n_s)\right]_{i,j}= \left\{
\begin{aligned}
N, && \kappa_i = \kappa_j\\
0, && \kappa_i \neq \kappa_j
\end{aligned}, 1\le i,j\le L.
\right.
\end{equation}
Note that since by design, $\kappa_i\neq\kappa_j,\forall i\neq j$, we have $\left[\bm{\Lambda}(n_s,n_s)\right]_{i,j}=0,\forall i\neq j$ and $\left[\bm{\Lambda}(n_s,n_s)\right]_{i,j}=N,\forall i= j$.
Thus the correlation matrix for $n_p=n_s$ is simply a scaled identity matrix, i.e., $\bm{\Lambda}(n_s,n_s)=N\mathbf{I}_L$.
Therefore, the output signal of MF in \eqref{pmf} reduces to
\vspace{-0.2cm}
\begin{equation}\label{mf3}
\vspace{-0.2cm}
{r}(n_s)=\alpha{\sqrt{N\mathbf{a}^H(\theta)\mathbf{F}\mathbf{F}^H\mathbf{a}(\theta)}}+\hat{{z}}.
\end{equation}
Therefore, the resulting SNR for DAM sensing is
\begin{equation}\label{sens_snr}
\begin{array}{l}
\gamma_p= \frac{|\alpha|^2 {{N\mathbf{a}^H(\theta)\mathbf{F}\mathbf{F}^H\mathbf{a}(\theta)}}}{\sigma^2}=\frac{|\alpha|^2 {{N\mathbf{a}^H(\theta)\left(\sum_{l=1}^L\mathbf{f}_l\mathbf{f}_l^H\right)\mathbf{a}(\theta)}}}{\sigma^2}.
\end{array}
\end{equation}
Note that with the average transmit power constraint in \eqref{power_const}, the maximum achievable SNR for sensing without considering the communication UE is
$\gamma_{p,\max}=\frac{|\alpha|^2NMP}{\sigma^2}$, which is achieved by letting $\mathbf{f}_l=\sqrt{\frac{P}{ML}}\mathbf{a}(\theta), \forall l=1,\cdots,L$.

\subsection{Beamforming Optimization for DAM-ISAC}
It is observed from \eqref{comm_snr} and \eqref{sens_snr} that both communication and sensing performance critically depend on the DAM beamforming vectors $\left\{\mathbf{f}_l\right\}_{l=1}^L$.
In this section, we consider the beamforming optimization problem to maximize the communication SNR in \eqref{comm_snr}, subject to the ISI-ZF condition while guaranteeing the sensing output SNR in \eqref{sens_snr} is no smaller than certain threshold.
By discarding the constant term $\sigma^2$ in \eqref{comm_snr}, the problem can be formulated as
\vspace{-0.2cm}
\begin{equation}\label{P1.1}
\vspace{-0.2cm}
\begin{aligned}
\mathrm{(P1)} &\max\limits_{\left\{\mathbf{f}_l\right\}_{l=1}^L}\ &&\left|\sum\nolimits_{l=1}^L\mathbf{h}_l^H\mathbf{f}_l\right|^2 \\
&\quad \text{s.t.}  &&\mathbf{h}_l^H\mathbf{f}_{l'}=0, \forall l\neq l',\\
& &&|\alpha|^2 {{N\mathbf{a}^H(\theta)\left(\sum\nolimits_{l=1}^L\mathbf{f}_l\mathbf{f}_l^H\right)\mathbf{a}(\theta)}}\ge\gamma_{th}\sigma^2, \notag\\
&      &&\sum\nolimits_{l=1}^L\left\|\mathbf{f}_l\right\|^2\le P,
\end{aligned}
\end{equation}
where $\gamma_{th}$ is the required SNR threshold to guarantee the desired sensing performance.

Denote by $\mathbf{H}_l=[\mathbf{h}_1,\cdots,\mathbf{h}_{l-1},\mathbf{h}_{l+1},\cdots,\mathbf{h}_L]\in\mathbb{C}^{M\times(L-1)},\forall l$.
Then the ISI-ZF constraint of $\mathrm{(P1)}$ can be equivalently expressed as $\mathbf{H}_l^H\mathbf{f}_l=\mathbf{0}_{(L-1)\times1},\forall l$, i.e., $\mathbf{f}_l$ should lie in the nullspace of $\mathbf{H}_l^H$.
Therefore, denote by $\mathbf{Q}_l\triangleq\mathbf{I}_M-\mathbf{H}_l(\mathbf{H}_l^H\mathbf{H}_l)^{-1}\mathbf{H}_l^H$ the project matrix into the space orthogonal to the columns of $\mathbf{H}_l$, then we have $\mathbf{f}_l = \mathbf{Q}_l\mathbf{b}_l,\forall l$, where $\mathbf{b}_l\in\mathbb{C}^{M\times1}$ denotes the new vector to be optimized.
As a result, problem $\mathrm{(P1)}$ can be further simplified as
\vspace{-0.2cm}
\begin{equation}\label{P2}
\vspace{-0.2cm}
\begin{aligned}
\mathrm{(P2)}\ &\max\limits_{\left\{\mathbf{b}_l\right\}_{l=1}^L}\ \left|\sum\nolimits_{l=1}^L\mathbf{h}_l^H\mathbf{Q}_l\mathbf{b}_l\right|^2 \\
&\quad\text{s.t.}\quad \sum\limits_{l=1}^L\mathbf{b}_l^H\mathbf{A}_l(\theta)\mathbf{b}_l\ge\tilde{\gamma}_{th}, \sum\nolimits_{l=1}^L\left\|\mathbf{Q}_l\mathbf{b}_l\right\|^2\le P, \notag
\end{aligned}
\end{equation}
where $\mathbf{A}_l(\theta)\triangleq\mathbf{Q}_l^H\mathbf{a}(\theta)\mathbf{a}^H(\theta)\mathbf{Q}_l, \forall l$, and $\tilde{\gamma}_{th}\triangleq\frac{\gamma_{th}\sigma^2}{|\alpha|^2N}$.
Note that due to the ISI-ZF constraint for communication, the optimal beamforming vectors can only be obtained in the nullspace of $\mathbf{H}_l^H$.
Therefore, without considering the communication performance, the maximum achievable SNR for sensing can be obtained when $\mathbf{f}_l=\frac{\sqrt{P}\mathbf{Q}_l\mathbf{a}(\theta)}{\sqrt{\sum\nolimits_{l=1}^L\left\|\mathbf{Q}_l\mathbf{a}(\theta)\right\|^2}}, l=1,\cdots,L$, and the resulting SNR is
\vspace{-0.2cm}
\begin{equation}
\vspace{-0.2cm}
\gamma_{\text{zf},\max}=\frac{|\alpha|^2NP\sum\nolimits_{l=1}^L\left\|\mathbf{a}^H(\theta)\mathbf{Q}_l\mathbf{a}(\theta)\right\|^2}{\sigma^2\sum\nolimits_{l=1}^L\left\|\mathbf{Q}_l\mathbf{a}(\theta)\right\|^2}\le\gamma_{p,\max}, \notag
\end{equation}
which is communication channel dependent.
Furthermore, since $\mathbf{Q}_l^H\mathbf{Q}_l=\mathbf{Q}_l$, the equality holds when $\mathbf{Q}_l=\zeta\mathbf{I}_M$, $\forall \zeta\neq 0$, $l=1,\cdots,L$.

By letting $\bar{\mathbf{b}}=\left[\mathbf{b}_1^H,\cdots,\mathbf{b}_L^H\right]^H\in\mathbb{C}^{ML\times1}$ and $\bar{\mathbf{h}}=\left[\mathbf{h}_1^H\mathbf{Q}_1,\cdots,\mathbf{h}_L^H\mathbf{Q}_L\right]^H\in\mathbb{C}^{ML\times1}$, problem $\mathrm{(P2)}$ can be further transformed to
\vspace{-0.2cm}
\begin{align}\label{P2.1}
\vspace{-0.2cm}
\mathrm{(P2.1)}\ &\max\limits_{\bar{\mathbf{b}}}\ \bar{\mathbf{b}}^H\bar{\mathbf{H}}\bar{\mathbf{b}} \notag\\
&\quad \text{s.t.} \ \bar{\mathbf{b}}^H\bar{\mathbf{A}}(\theta)\bar{\mathbf{b}}\ge\tilde{\gamma}_{th}, \\
&\quad\quad\ \ \bar{\mathbf{b}}^H\bar{\mathbf{Q}}\bar{\mathbf{b}}\le P,
\end{align}
where $\bar{\mathbf{H}}\triangleq\bar{\mathbf{h}}\bar{\mathbf{h}}^H$, $\bar{\mathbf{A}}(\theta)\triangleq\mathrm{diag}\left(\mathbf{A}_1(\theta),\cdots,\mathbf{A}_L(\theta)\right)$, and $\bar{\mathbf{Q}}\triangleq\mathrm{diag}\left(\mathbf{Q}_1,\cdots,\mathbf{Q}_L\right)\in\mathbb{C}^{ML\times ML}$.
Note that the objective function and the left hand side of constraint (21)  are nonconcave, thus $\mathrm{(P2.1)}$ is a nonconvex quadratic constraint quadratic programming (QCQP), which cannot be directly solved by the standard convex optimization techniques.

To find the efficient suboptimal solution to $\mathrm{(P2.1)}$, we utilize successive convex approximation (SCA) technique \cite{zeng2019accessing}, which is an iterative optimization method based on global convex lower bounds.
Specifically, for each iteration $i$, the objective function of $\mathrm{(P2.1)}$ can be globally lower bounded as
\vspace{-0.2cm}
\begin{equation}\label{lbo}
\vspace{-0.2cm}
\begin{aligned}
\bar{\mathbf{b}}^H\bar{\mathbf{H}}\bar{\mathbf{b}}&\ge\bar{\mathbf{b}}_{i}{^H}\bar{\mathbf{H}}\bar{\mathbf{b}}_{i} \\
&+2\Re\left\{\bar{\mathbf{b}}_{i}^{H}\bar{\mathbf{H}}\left(\bar{\mathbf{b}}-\bar{\mathbf{b}}_{i}\right)\right\}\triangleq\varphi_\mathrm{lb}^{(i)}(\bar{\mathbf{b}}), \forall \bar{\mathbf{b}},\\
\end{aligned}
\end{equation}
where $\bar{\mathbf{b}}_{i}$ is a given local point at the $i$th iteration.
\begin{algorithm}[t]
\caption{SCA for Solving $\mathrm{(P2.1)}$} \label{alg}
\hspace*{0.02in} {\bf Input:} 
$\left\{\mathbf{h}_l\right\}_{l=1}^L$, $\theta$, $\tilde{\gamma}_{th}$, $P$\\
\hspace*{0.02in} {\bf Output:} 
$\bar{\mathbf{b}}$
\begin{algorithmic}[1]
\State Initialize $\bar{\mathbf{b}}_{0}$. Let $i=0$. 
\Repeat
\State Obtain $\varphi_\mathrm{lb}^{(i)}(\bar{\mathbf{b}})$ and $\psi_\mathrm{lb}^{(i)}(\bar{\mathbf{b}})$ in problem $\mathrm{(P2.2)}$, for the given local point $\bar{\mathbf{b}}_{i}$.
\State Solve problem $\mathrm{(P2.2)}$ to get the optimal solution $\bar{\mathbf{b}}_{i}^{\star}$
\State Update the local point $\bar{\mathbf{b}}_{i+1}=\bar{\mathbf{b}}_{i}^{\star}$
\State Update $i=i+1$
\Until{The increase of the objective value of $\mathrm{(P2.1)}$ below a threshold $\epsilon$.}
\end{algorithmic}
\end{algorithm}
Note that \eqref{lbo} follows from the fact that $\bar{\mathbf{b}}^H\bar{\mathbf{H}}\bar{\mathbf{b}}$ is a convex function, and the first-order Taylor expansion of a convex differentiable function is its global under-estimator \cite{boyd2004convex}.
Furthermore, the equality of \eqref{lbo} holds at the point $\bar{\mathbf{b}}=\bar{\mathbf{b}}_{i}$, and the objective function and its lower bound $\varphi_\mathrm{lb}^{(i)}(\bar{\mathbf{b}})$ have identical gradient, which is equal to $2\bar{\mathbf{H}}\bar{\mathbf{b}}_{i}$.
Similarly, we have the lower bound for the left hand side of (21) as
\vspace{-0.15cm}
\begin{equation}\label{lbs}
\vspace{-0.15cm}
\begin{aligned}
\bar{\mathbf{b}}^H\bar{\mathbf{A}}(\theta)\bar{\mathbf{b}}&\ge\bar{\mathbf{b}}_{i}^{H}\bar{\mathbf{A}}(\theta)\bar{\mathbf{b}}_{i} \\
&+2\Re\left\{\bar{\mathbf{b}}_{i}^{H}\bar{\mathbf{A}}(\theta)\left(\bar{\mathbf{b}}-\bar{\mathbf{b}}_{i}\right)\right\}\triangleq\psi_{\mathrm{lb}}^{(i)}(\bar{\mathbf{b}}), \forall \bar{\mathbf{b}}.
\end{aligned}
\end{equation}
Then, by replacing the nonconcave objective functions and the constraint (21) in $\mathrm{(P2.1)}$ with their corresponding concave lower bounds, we have the following convex optimization problem
\vspace{-0.2cm}
\begin{align}\label{P2.2}
\vspace{-0.2cm}
\mathrm{(P2.2)}\ &\max\limits_{\bar{\mathbf{b}}}\ &&\varphi_\mathrm{lb}^{(i)}(\bar{\mathbf{b}}) \notag\\
&\text{s.t.} &&\psi_{\mathrm{lb}}^{(i)}(\bar{\mathbf{b}})\ge\tilde{\gamma}_{th}, \bar{\mathbf{b}}^H\bar{\mathbf{Q}}\bar{\mathbf{b}}\le P.
\end{align}
Problem $\mathrm{(P2.2)}$ can be efficiently solved by using the standard convex optimization techniques or convex optimization toolbox, like CVX.
Note that the new constraint $\psi_{\mathrm{lb}}^{(i)}(\bar{\mathbf{b}})\ge\tilde{\gamma}_{th}$ always implies the constraint (21) in $\mathrm{(P2.1)}$.
Therefore, the objective value of $\mathrm{(P2.2)}$ gives at least a lower bound to that of problem $\mathrm{(P2.1)}$.
Thus, by iteratively updating the local point with $\bar{\mathbf{b}}_{i+1}=\bar{\mathbf{b}}_{i}^{\star}$, and solving a sequence of convex optimization problem $\mathrm{(P2.2)}$, an effective solution of the original nonconvex problem $\mathrm{(P2.1)}$ and hence that of problem $\mathrm{(P1)}$ can be obtained, which is summarized in Algorithm~\ref{alg}.

\vspace{-0.1cm}
\section{DAM Versus OFDM For Sensing}
\begin{figure} 
  \setlength{\abovecaptionskip}{-0.1cm}
  \setlength{\belowcaptionskip}{-0.3cm}
  \centering
  \subfigure[DAM block structure]{
  \includegraphics[width=0.38\textwidth]{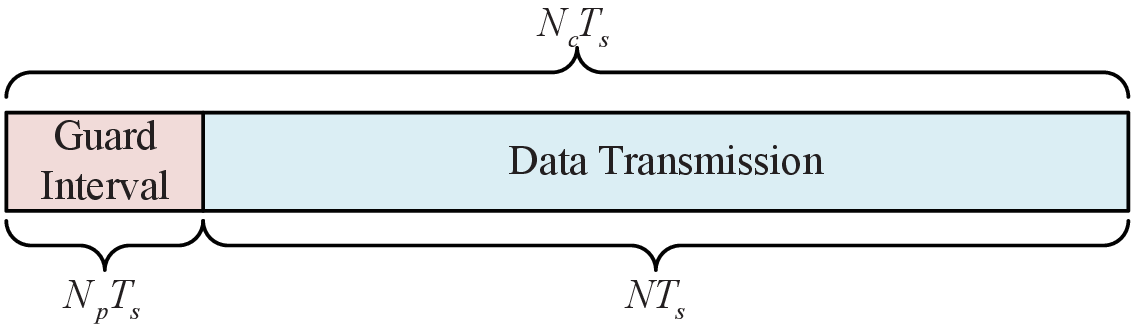}\label{damb}
  }
  \\
  \subfigure[OFDM block structure]{
  \includegraphics[width=0.38\textwidth]{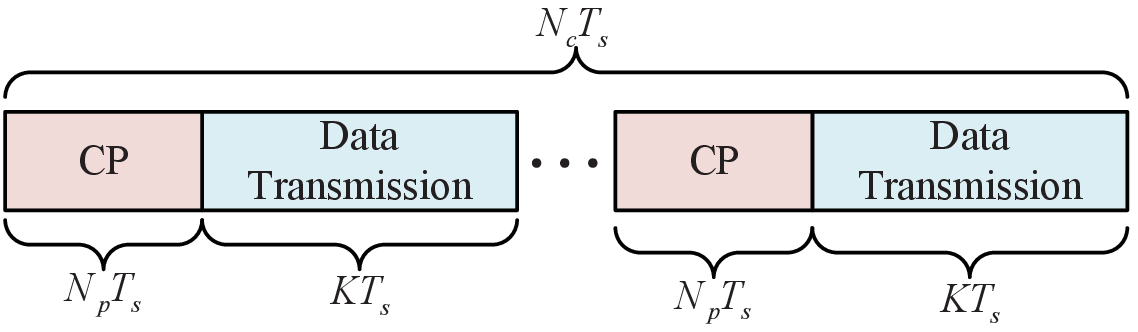}\label{ofdm}
  }
  \caption{An illustration of DAM and OFDM block structures \cite{dam2021lu}.}\label{block}\vspace{-10pt}
\end{figure}

\vspace{-0.1cm}
In this section, we compare the sensing performance between DAM and OFDM, while their communication performance has been compared in \cite{dam2021lu}.
Denote by $T_c$ the channel coherent time, during which the communication and sensing channels are assumed to be unchanged.
Therefore, the length of a transmission block is $N_c=T_c/T_s$, with $T_s=1/B$.
Furthermore, denote by $T_p=N_pT_s$ the guard interval, which is no smaller than the maximum delay over all channel coherence blocks, i.e., $N_p\ge \tilde{n}_{\max}$, with $\tilde{n}_{\max} \ge n_{\max}$.
Therefore, as shown in Fig.~\ref{damb}, we have the block structure of DAM as $N_c=N+N_p$, where $N$ denotes the number of DAM symbols for each block, which is also the number of DAM symbols per CPI for sensing in \eqref{dam_tx2}.
On the other hand, for OFDM with $K$ subcarriers, the subcarrier spacing is $\triangle f = \frac{B}{K} = \frac{1}{KT_s}$ and OFDM symbol duration is $T=\frac{1}{\triangle f} = K T_s$.
Assuming the cyclic prefix (CP) of length $N_p$, we have the OFDM block structure as shown in Fig.~\ref{ofdm}, with $N_c=I(K+N_p)$, where $I$ denotes the number of OFDM symbols for each block.
Note that as illustrated in Fig.~\ref{block}, we have $N>IK$ thanks to the saving of guard interval of DAM.

\vspace{-0.2cm}
\subsection{Ambiguity and Resolution Analysis}
\vspace{-0.2cm}
Denote by $x_{k,i}$ the $i$th transmitted OFDM symbol at the $k$th subcarrier, with $\mathbb{E}[|x_{k,i}|^2] = 1$, where $k=1,\cdots,K$, and $i=1,\cdots,I$.
For MISO-OFDM radar, the $i$th received OFDM symbol at the $k$th subcarrier can be expressed as
\vspace{-0.15cm}
\begin{equation}\label{ofdm_rx}
\vspace{-0.15cm}
y_r^{(k,i)} = \alpha \mathbf{a}^H(\theta)\mathbf{w}_kx_{k,i}e^{j2\pi iT_of_d}e^{-j2\pi k\triangle f \tau} + z_{k,i},
\end{equation}
where $T_o=(N_p+K)T_s$ is the total OFDM symbol duration including the CP, and $\mathbf{w}_k,k=1,\cdots,K$ denotes the transmit beamforming vector of the $k$ subcarriers, with $\left\|\mathbf{w}_k\right\|^2\le P_k$, where $P_k$ is the transmit power of the $k$th subcarriers.
It is observed from \eqref{ofdm_rx} that, for OFDM radar to estimate the delay $\tau$ and Doppler frequency $f_d$, the orthogonality of the received modulation symbols in time and frequency domain is required.
Specifically, to preserve the orthogonality in time domain, the maximum delay should be no greater than the length of CP, i.e., $0\le \tau\le N_pT_s$ \cite{sturm2011waveform}.
In frequency domain, the maximum tolerable Doppler frequency shift is only $10\%$ of the subcarrier spacing, i.e., $f_{d}\le \triangle f/10$.
Therefore, the maximum unambiguity range and velocity of OFDM radar are only $R_{\max} = \frac{cN_pT_s}{2}$ and $\left|v_{d,\max}\right|=\frac{\lambda}{20KT_s}$, respectively, with the resolution of $\triangle R =\frac{c}{2B}$ and $\triangle v = \frac{\lambda}{2N_cT_s}$.

On the other hand, for the proposed DAM, in each transmission block, the first $N_p$ elements corresponding to the guard interval are discarded, thus the delay and Doppler frequency of the target can be estimated based on the remained signals as discussed in Section~\ref{performance analysis}.
However, if delay $n_s$ exceeds the length of guard interval $N_p$, there will be ISI across different blocks.
Thus, the unambiguity delay for DAM is $0\le n_s\le N_p$, which is the same as that of the OFDM.
However, as a SC waveform, the unambiguity Doppler frequency of DAM is $\left|f_d\right|\le \frac{1}{2T_s}$, which corresponds to the maximum unambiguity velocity of $\left|v_{d,\max}\right| =\frac{\lambda}{2T_s}$.
Therefore, compared to OFDM, DAM outperforms OFDM in terms of the maximum unambiguity velocity.

\subsection{PAPR Analysis}
For the considered MISO-OFDM radar, after signal processing, for the particular delay-Doppler bin where the target actually lies, the output SNR can be obtained as
\vspace{-0.2cm}
\begin{equation}
\vspace{-0.2cm}
\gamma_{\text{OFDM}} = \frac{|\alpha|^2I\sum\nolimits_{k=1}^K\left\|\mathbf{a}^H(\theta)\mathbf{w}_k\right\|^2}{\sigma^2/K},
\end{equation}
where $I$ corresponds the OFDM radar processing gain over a block of $I$ OFDM symbols.
Thus, with the target direction $\theta$ known, it is not difficult to see that the maximum output SNR is $\gamma_{\text{OFDM},\max}=\frac{|\alpha|^2MIK\sum\nolimits_{k=1}^LP_k}{\sigma^2}$, when $\mathbf{w}_k=\sqrt{\frac{P_k}{M}}\mathbf{a}(\theta)$.
On the other hand, for DAM over a block, the the maximum output SNR is $\gamma_{p,\max} = \frac{\left|\alpha\right|^2MNP}{\sigma^2}$.
Here, for fair comparison, it is assumed that $P=\sum\nolimits_{k=1}^KP_k$, thus we have $\gamma_{p,\max}>\gamma_{\text{OFDM},\max}$, since $N>KI$.
Therefore, DAM can accumulate more energy of the echo signal for sensing, thanks to its significantly reduced guard interval overhead.
Moreover, OFDM radar suffers from high PAPR, which further limits its sensing capability.
Specifically, taking the $M$-phase shift keying (PSK) transmission as an example, for $M$-PSK OFDM with $K$ subcarriers, the resulting PAPR is $\eta_{\text{OFDM}} = K$.
However, for the proposed DAM, the maximum transmit power occurs when all $L$ paths signals add coherently, thus the resulting PAPR is $\eta_{\text{DAM}}= L$.
Therefore, when the maximum transmit power is fixed to $P_{\max}$, the maximum output SNR of OFDM radar is
$\gamma_{\text{OFDM},\max} = \frac{\left|\alpha\right|^2MIKP_{\max}}{\eta_{\text{OFDM}}\sigma^2} = \frac{\left|\alpha\right|^2MIP_{\max}}{\sigma^2}$, while that of DAM is
\vspace{-0.2cm}
\begin{equation}
\vspace{-0.2cm}
\gamma_{p,\max} = \frac{\left|\alpha\right|^2MNP_{\max}}{\eta_{\text{DAM}}\sigma^2}= \frac{\left|\alpha\right|^2MNP_{\max}}{L\sigma^2}\gg \gamma_{\text{OFDM},\max}, \notag
\end{equation}
which is much greater than that of OFDM since we typically have $K\gg L$ and thus $N/L\gg N/K>I$.
\vspace{-0.2cm}
\section{Simulation Results}
\begin{figure} 
  \setlength{\abovecaptionskip}{-0.1cm}
  \setlength{\belowcaptionskip}{-0.3cm}
  \centering
  \includegraphics[width=0.4\textwidth]{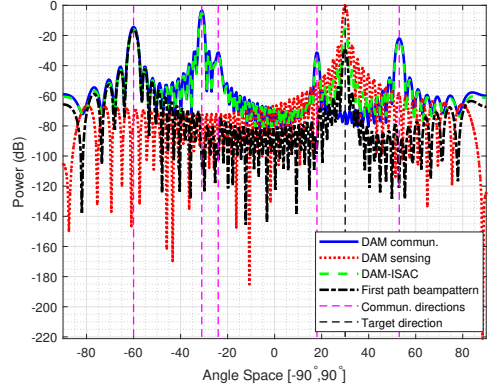}
  \caption{Transmit beampatterns for DAM ISI-ZF communication, DAM radar, and DAM ISAC beamformers, where the directions of the communication multi-paths are $[-60^\circ,-31^\circ,-24^\circ,18^\circ,54^\circ]$, the sensing target direction is $\theta = 30^\circ$, and the number of temporal-resolvable paths is $L=5$.}\label{beampattern}\vspace{-0.6cm}
\end{figure}
In this section, simulation results are provided to evaluate the performance of the proposed DAM technique for ISAC.
The total bandwidth is $B=100$ MHz, and the carrier frequency is $f_c = 28$ GHz, which corresponds the temporal resolution  $T_s = 10$ ns and the range resolution  $\triangle R =1.5$ m.
The channel coherent time is $T_c=1$ ms, thus the length of a transmission block is $N_c=1\times10^5$.
The guard interval (or CP) is $T_p = 2\ \upmu$s, which corresponds the maximum unambiguity range  $R_{\text{ua}} = 300$~m, $N_p=T_g/T_s=200$, and the number of DAM symbols is $N = N_c-N_p=9.98\times 10^4$.
The transmit power is $P = 30$ dBm, and the transmitter is equipped with an uniform linear array (ULA) consisting of $M=64$ transmitting antennas.
The noise power spectral density is $-169$ dBm/Hz.
For radar sensing, the two-way channel gain between the ISAC transmitter and target is modelled as $|\alpha|^2=\frac{\lambda^2\xi}{(4\pi)^3R^4}$, where $R=200$ m is the sensing distance and $\xi=1$ $\text{m}^2$ is the radar cross section of the target.
Similar to \cite{dam2021lu}, for mmWave communication, the channel of each delay path is modelled as $\mathbf{h}_l=\beta_l\sum\nolimits_{i=1}^{\mu_{l}}v_{li}\mathbf{a}(\theta_{li})$, where $\beta_l$ denotes the complex channel coefficient for the $l$th path, while $\mu_l$ denotes the number of subpaths for the $l$th path with same delay but different AoDs $\theta_{li},i=1,\cdots, \mu_l$; $v_{li}$ is the complex coefficient of the $i$th subpath for the $l$th path.
Here, it is assumed that $\mu_l$ is uniformly distributed in $[1,\mu_{\max}]$, with $\mu_{\max}=3$, while the feasible AoDs within $[-60^{\circ},60^{\circ}].$

Fig.~\ref{beampattern} shows the transmit beampattern for different transmit beamforming schemes.
It is observed that for ISI-ZF communication without considering sensing, DAM beamforming only forms beams towards the directions of the multi-paths of the communication channel, i.e., $[-60^\circ,-31^\circ,-24^\circ,18^\circ,54^\circ]$.
On the other hand, if the radar sensing is only considered, the formed beam only points towards the sensing target direction $\theta=30^\circ$.
By contrast, when both communication and sensing are considered, the DAM-ISAC beamforming successfully generates 6 beams towards all possible directions for both communication and sensing.
Thus, it can successfully track the target while providing ISI-free communication for the UE.
To further illustrate this point, the DAM-ISAC beampattern for the first multipath, i.e., $\left\|\mathbf{f}_1^H\mathbf{a}(\theta_t)\right\|^2,\forall \theta_t\in[-90^\circ,90^\circ]$, is also shown in Fig.~\ref{beampattern} by the black dot line, where the resulting beam only points towards the direction of the first multipath and the target, while suppressed towards other multipath directions to enable ISI-free communication.

\begin{figure} 
  \setlength{\abovecaptionskip}{-0.1cm}
  \setlength{\belowcaptionskip}{-0.3cm}
  \centering
  \includegraphics[width=0.4\textwidth]{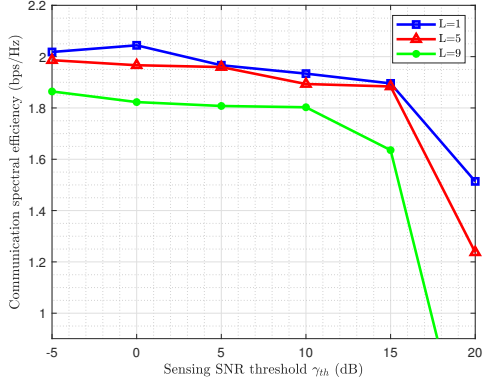}
  \caption{Average communication spectral efficiency versus sensing SNR threshold $\gamma_{th}$ for DAM-ISAC with ISI-ZF beamforming.}\label{rateVersusSNR}\vspace{-0.6cm}
\end{figure}
Fig.~\ref{rateVersusSNR} shows the average communication spectral efficiency in bps/Hz over $10^3$ channel realization versus the sensing SNR threshold $\gamma_{th}$ for DAM-ISAC with ISI-ZF beamforming.
For communication, the spectral efficiency over a transmission block is given by $C_{\text{DAM}}=\frac{N}{N_c}\log_2\left(1+\gamma_c^\star\right)$ \cite{dam2021lu}, where $\gamma_c^\star$ is the output SNR obtained by solving the optimization problem $(\mathrm{P2.1})$.
It is observed that the communication spectral efficiency degrades as the required sensing SNR increases.
This is expected since as the sensing SNR threshold increases, the beams should be formed to better match the target direction, resulting the reduction of $\gamma_c^*$ and hence the communication spectral efficiency.
On the other hand, it is observed that the communication spectral efficiency reduces as $L$ increases, since DAM is most effective when the number of antennas $M$ is considerably larger than the number of delay paths $L$, in which case the propagation delay of each path can be pre-compensated independently without affecting other paths.

\section{Conclusion}
This paper studied ISAC with the novel DAM technique, which is a low-complexity SC transmission scheme that exploits the high spatial dimension and multi-path sparsity of mmWave/Terhertz massive MIMO channel.
On one hand, DAM enables ISI-free communication without requiring sophisticated channel equalization.
On the other hand, such a SC transmission scheme has low PAPR, which is expected to outperform OFDM in terms of maximum unambiguity velocity and output sensing SNR.
Simulation results demonstrated that the proposed DAM-ISAC beamforming method can simultaneously provide the ISI-free communication with high spectral efficiency while tracking the target for sensing.

%

\section*{Acknowledgment}
This work was supported by the National Key R\&D Program of China with grant number 2019YFB1803400.

\bibliographystyle{IEEEtran}
\bibliography{DAM}

\end{document}